\documentclass[aps,prl,twocolumn,superscriptaddress,showpacs]{revtex4}

\usepackage{longtable}
\usepackage{graphicx}
\usepackage{dcolumn}
\usepackage{bm}
\begin{document}


\title{Dirty Superconductivity in the Electron-Doped Cuprate Pr$_{2-x}$Ce$_x$CuO$_{4-\delta}$:\\ a Tunneling Study}


\author{Y. Dagan}
\email[]{yodagan@post.tau.ac.il} \affiliation{School of Physics
and Astronomy, Raymond and Beverly Sackler Faculty of Exact
Sciences, Tel-Aviv University, Tel Aviv, 69978, Israel}
\affiliation{Center for Superconductivity Research Physics
Department University of Maryland College Park, MD, 20743}
\author{R. Beck}
\affiliation{School of Physics and Astronomy, Raymond and Beverly
Sackler Faculty of Exact Sciences, Tel-Aviv University, Tel Aviv,
69978, Israel}
\author{R.L. Greene}
\affiliation{Center for Superconductivity Research Physics
Department University of Maryland College Park, MD, 20743}


\date{\today}

\begin{abstract}
We report a tunneling study between
Pr$_{2-x}$Ce$_x$CuO$_{4-\delta}$ and Lead as a function of doping,
temperature and magnetic field. The temperature dependence of the
gap follows the BCS prediction. Our data fits a nonmonotonic
$d$-wave order parameter for the whole doping range studied. From
our data we are able to conclude that the electron-doped cuprate
Pr$_{2-x}$Ce$_x$CuO$_{4-\delta}$ is a weak coupling, BCS, dirty
superconductor.
\end{abstract}


\maketitle

In the theory for superconductivity by Bardeen, Cooper and
Shrieffer (BCS) as the temperature is raised from T=0 K the
amplitude of the superconducting order-parameter decreases and
eventually becomes zero at the critical temperature T$_c$
\cite{BCS}. This is not necessarily the case for the high T$_c$
cuprate superconductors. For example scanning tunneling
spectroscopy showed a non vanishing gap above T$_c$ and a constant
gap amplitude below T$_c$ for hole doped cuprates
\cite{Rennerandfisher}. It has been proposed \cite{Emerykivelson}
that in the high T$_c$ cuprates the amplitude of the
order-parameter does not go to zero at T$_c$ but phase
fluctuations eventually destroy long range coherence. The Uemura
plot \cite{Uemuraplot}, showing that T$_c$ scales with the
superfluid density gave further support to this scenario. The
electron-doped cuprates, however, fall off the Uemura line
\cite{opticalpropertiesNCCO}and this raises the fundamental
question whether the order-parameter falls to zero at T$_c$ for
these compounds.
\par
The symmetry of the order-parameter in the electron-doped cuprates
is still a matter of debate. While many experiments suggest an
order-parameter having a $d$-wave symmetry
\cite{TsueiElectronDoped,Prozorov,BalciPhysRevB,Hilgenkamp} others
suggest a change of symmetry with doping
\cite{Biswas,LembergerSkinta}. Raman spectroscopy on optimally
doped samples has been interpreted in terms of non-monotonic
$d$-wave (nm$d$) in which higher harmonics have significant
contribution \cite{blumbergNmd}. In this case the amplitude of the
order-parameter has a maximum at angle smaller than 45$^o$ to the
nodal direction. Further support for this scenario was later found
experimentally from angle resolved photoemission spectroscopy
\cite{matsui:017003} and from theoretical calculations
\cite{krotkov:107002}. Yet, it is still a mystery why the
tunneling spectra for electron-doped cuprates are isotropic in the
$ab$ plane \cite{shan:144506} and missing the expected zero bias
conductance peak in low transparency junctions \cite{Tanakafirst}.
Furthermore, for such order-parameter any scattering center is a
pair breaker that destroys superconductivity in its vicinity
\cite{VarmaSachdevMillis}. For hole doped cuprates the coherence
length, $\xi_{BCS}$ is usually much shorter than the mean free
path $\ell$, thus allowing the order-parameter to recover between
scattering events. For electron-doped cuprates $\xi_{BCS}$ is an
order of magnitude larger than for the hole doped. It is therefore
important to find out whether these compounds are in the clean
$(\xi_{BCS}\ll\ell)$ or dirty regime.
\par
In this letter we report a tunneling study as a function of
doping, temperature, and magnetic field using
lead/Insulator/Pr$_{2-x}$Ce$_x$CuO$_{4-\delta}$ (Pb/I/PCCO) planar
junctions. We find that the order-parameter goes to zero in a BCS
like fashion. This rules out phase fluctuations as the reason for
loss of coherence at T$_c$. We obtain a good fit to a
non-monotonic $d$-wave order-parameter in the PCCO electrode for
the whole doping (up to x=0.19) and temperature range studied.
From our tunneling data we are able to conclude that PCCO is in
the dirty limit. This suggests a possible solution for the long
standing puzzles of tunneling isotropy and the absence of zero
bias conductance peak in electron-doped cuprates.
\begin{figure}
 \includegraphics[width=0.95\hsize]{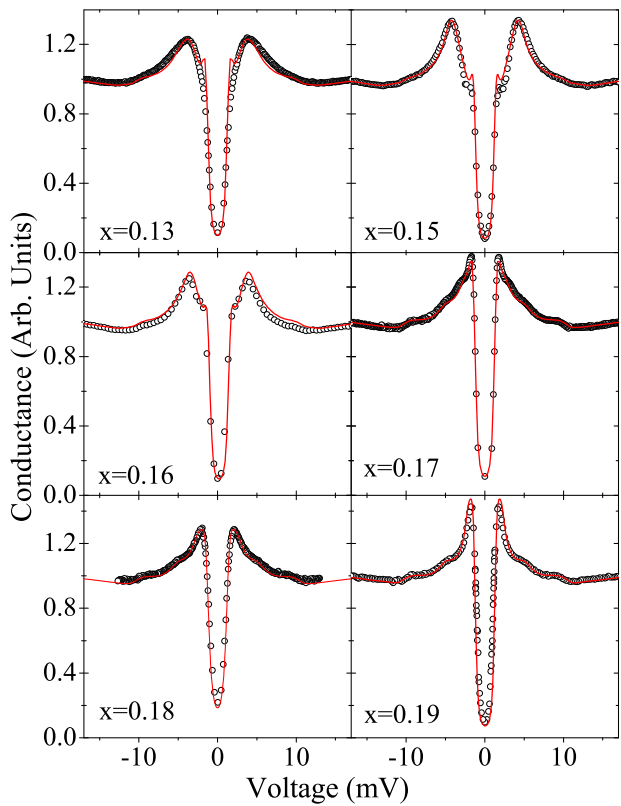}
\caption{(color online) Conductance versus voltage for the six
doping levels of Pr$_{2-x}$Ce$_x$CuO$_{4-\delta}$ studied at 1.8K
and zero field (circles). At this temperature and field both
electrodes are superconducting: note the Pb phonons around 5 and 9
meV. The red solid line is a non-monotonic $d$-wave fit to the
data described in the text.   \label{MainResults}}
 \end{figure}
\par
In a tunneling experiment a quasiparticle is injected into a
superconductor through a dielectric barrier. This method has been
proven to be a powerful tool for probing the density of states in
conventional superconductors \cite{GiaeverRMP}. A theory was later
developed to account for barriers of various transparencies using
a single parameter Z \cite{BTK}. It was later extended for the
case of anisotropic materials and order-parameters
\cite{Tanakafirst}.
\par
Lead counter electrodes were deposited on PCCO $c-$axis oriented
films using a method described elsewhere
\cite{daganQazilbashtunneling}. This procedure results in good
tunnel junctions with a lead-oxide barrier \cite{naitobarrier}. At
zero field and below T$_c$(Pb), we obtain
superconductor/I/superconductor (SIS) tunneling spectra. These
spectra exhibit conductance close to zero at low biases and Pb
phonon signatures, indicating good tunnel
junctions\cite{GiaeverRMP}. This is different than grain boundary
junctions where the leakage current was reported to depend on the
doping level \cite{Alffnat}. Previously, we showed that the
spectra obtained on $ab$ faces of single crystals are identical to
those of $c-$axis oriented films. This suggests dominating in
plane tunneling for the latter case\cite{daganQazilbashtunneling}.
The presence of in-plane nano-facets and a much reduced tunneling
probability in the $c$ direction \cite{daganEPJB} can be the
reason for the in-plane dominance. Upon applying a 14T magnetic
field perpendicular to the $ab$ planes and to the junction, one
drives both electrodes into their normal (N) state. This allows us
to normalize each spectrum with its respective 14T one, thus
cancelling extrinsic effects coming from either the junction or
the counter electrode. This is the standard procedure used in
classical tunneling experiments\cite{GiaeverRMP}. This procedure
was repeated for magnetic fields up to 14T and temperatures up to
30K. We define T$_c$(PCCO) at the junction as the temperature at
which the zero field spectrum merges with its corresponding 14T
one. The upper critical field, H$_{c2}$, at 1.8K is the field at
which the spectrum merges with that of the 14T one.
\par
\begin{table*}
\caption{Superconducting parameters for the various junctions. The
critical temperature T$_c$ and field H$_{c2}$ are measured using
the tunneling conductance as described in the text. All other
parameters are found at 1.8K.  $\Delta_{max}$ is the gap amplitude
reached at an angle $\theta_{max}$ measured from the a-axis.
$\Delta_{max}, \theta_{max}$ are found using the fitting function,
described in the text. For a simple $d$-wave order-parameter
$\theta_{max}=0$. $\xi_{GL}$ and $\xi_{BCS}$ are the
Ginzbureg-Landau and the BCS coherence lengths respectively.
$\ell$ can be calculated from the latter two length scales.
\label{myTable}}
 \begin{ruledtabular}
 \begin{tabular}{lcccccccccc}
   \hline
  Ce doping  & $T_c (K)$ & $\mu_0H_{c2} (Tesla)$ & $\Delta_{max} (meV)$ & $\theta_{max}$(deg) & Z & $\Gamma (meV)$ & $\frac{2\Delta_{max}}{kT_c}$ & $\xi_{GL} (nm)$ & $\xi_{BCS} (nm)$ & $\ell$ (nm)\\
  \hline
  0.13 &  17$\pm 1$  & 11$\pm 1$  & 2.5 & 16 & 4 & 1.3 & 3.5$\pm 0.5$  & 5.5$\pm 0.3$   & 31$\pm 9$  & 1.3$\pm 0.4$ \\
  0.15 &  19$\pm 1$  & 7$\pm 1$  & 3.25 & 18 & 4 & 0.9 & 4.0$\pm 0.4$  & 6.9$\pm 0.5$ &  24$\pm 6$ &  2.7 $\pm 0.8$\\
  0.16 &  16$\pm 1$  & 5$\pm 1$  & 2.6 & 18 & 4 & 0.95 & 3.8$\pm 0.5$  & 8.1$\pm 0.8$ &  30$\pm 8$ & 3 $\pm 1$ \\
  0.17 &  13$\pm 1$  & 3$\pm 1$  & 1.3 & 22 &  4.1 & 0.93 &2.3$\pm 1.3$  &  10.5$\pm 1.7$ &  61$\pm 21$ & 2.5 $\pm 1.2$ \\
  0.18 &  11$\pm 1$  & 1$\pm 0.2$ & 1.0 & 22 & 7 & 0.75\footnotemark[1]    & 2.1$\pm 1.5$  &  18$\pm 2$  & 79$\pm 31$  & 5.7$\pm 2.5$ \\
  0.19 &  8$\pm 0.4$ & 0.3$\pm 0.1$  & 0.9 & 23 & 4.9  & 0.45 & 2.6$\pm 0.9$  &   33$\pm 6$  & 87$\pm$ 36 & 17$\pm 9$  \\
  \hline
  \footnotetext[1]{An additional life-time broadening of 0.25meV was added to the Pb counter electrode.}

\end{tabular}
\end{ruledtabular}
\end{table*}
\par
In Fig.\ref{MainResults} we show the normalized S/I/S spectra
(circles) at T=1.8K and H=0T for doping levels ranging from
underdoping (x=0.13) through optimum doping (x=0.15) to heavy
overdoping (x=0.19). To fit the data a tunneling conductance,
$G(eV)$, can be calculated as follows:
\begin{eqnarray}
    \frac{G(eV)}{G_n}=\frac{d}{d(V)}\int&&D_1(E,T)D_2(E+eV,T) \times \nonumber\\
&& [f(E,T)-f(E+eV,T)]dE
\end{eqnarray}
where, $G_n$ is the conductance when both electrodes are in the
normal states, $V$ is the bias across the junction, $f$ is the
Fermi distribution function at a given temperature, $T$. $D_{1}$
is the tunneling density of states calculated as in Ref.\cite{BTK}
for the well known Pb density of states including the phonon
spectrum. $D_2$ is tunneling density of states calculated for PCCO
using the theory for anisotropic order-parameter in
Ref.\cite{Tanakafirst} after integrating over all angles using the
fitting parameters: $Z$, $\Delta_0, \eta$, and $\Gamma$, where Z
is the barrier strength, $\Gamma$ is a life-time
broadening\cite{dyneslifetime}. We introduce the following form
for a nm$d$ order-parameter:
\begin{equation} \label{eq:Delta}
\Delta(\theta)=\Delta_0\cos(2\theta)\{1+\frac{\eta\cos^2(2\theta)}{1-0.9|\cos(2\theta)|}\}^{-\frac{3}{2}}
\end{equation}
where $\theta$ is the angle between quasi-particle
momentum and the (1,0,0) direction. This function keeps the nodes
and phases as in the case of pure $d$-wave but shifts the gap
maximum towards the node with increasing $\eta$. The solid line in
Fig. \ref{MainResults} is the best fit obtained with the
parameters described in table \ref{myTable}. The gap amplitude
$\Delta_{max}$ and the angle from the 100 direction at which the
maximum gap is obtained $\theta_{max}$ are in agreement with
photoemission measurements\cite{matsui:017003}. Such an
order-parameter fits the data better than a pure $d$-wave, much
better than $s$-wave or any other linear combination of the two.
\par
\begin{figure}
 \includegraphics[width=1\hsize]{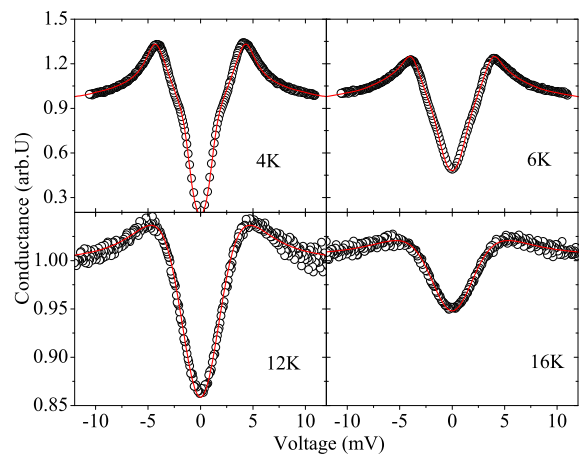}
 \caption {(color online) Lead / Pr$_{1.85}$Ce$_{0.15}$CuO$_{4-\delta}$ tunneling
 spectra at representative temperatures (circles). The line is a
 fit using the parameters obtained from Fig 1. with the gap
 amplitude ($\Delta_0(T)$) being the only fitting parameter.}

\label{TempGraph}
 \end{figure}
\begin{figure}

\includegraphics[width=1\hsize]{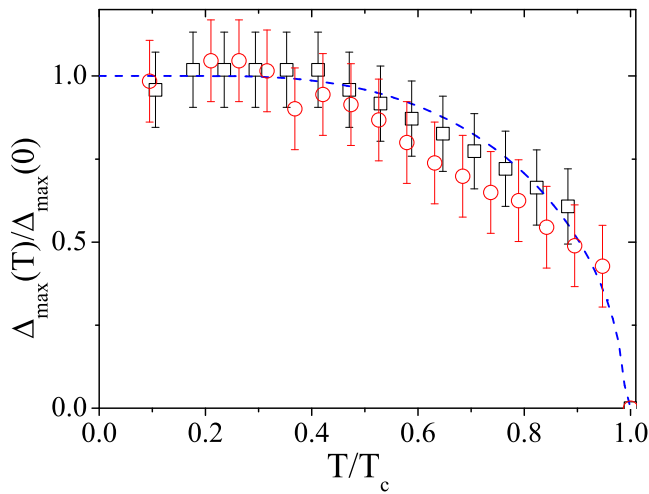}
\caption{(color online) The reduced gap maximum obtained from fits
as demonstrated in Fig.\ref{TempGraph} plotted \emph{versus}
reduced temperature for
Pr$_{1.87}$Ce$_{0.13}$CuO$_{4-\delta}$(squares) and
Pr$_{1.85}$Ce$_{0.15}$CuO$_{4-\delta}$ (circles). The dashed line
is the BCS universal line. \label{DeltaOfTgraph}}

 \end{figure}

For x=0.13 and x=0.15 we fit the tunneling characteristics up to
T$_c$. We use a BCS temperature dependence for the Pb
order-parameter and the measured temperature T. We keep all other
parameters used for 1.8K, constant varying only a single parameter
$\Delta_0$. In Fig.\ref{TempGraph} we show examples for such fits
below and above T$_c$(Pb)=7.2 K. The remarkable agreement between
the data and the fits give further support to the fitting
parameters used at the base temperature since they fit both the
S/I/S and the N/I/S cases. For the latter the calculation simply
reduced to that of Ref.\cite{Tanakafirst}. We follow the
temperature dependence of $\Delta_0$ obtained from the fits. This
allows us to find $\Delta_{max}(T)$ from Eq. \ref{eq:Delta}. The
obtained $\Delta_{max}(T)$ for x=0.13 (squares), x=0.15 (circles)
are plotted on the universal BCS graph in Fig.\ref{DeltaOfTgraph}.
We note a very good agreement to the BCS prediction (dashed
line)\cite{BCS}.
\par
The agreement with the BCS temperature dependence is in contrast
with the hole doped cuprate data \cite{Rennerandfisher}. This
behavior rules out phase fluctuations as the cause for the
disappearance of long range order superconductivity at $T_c$. This
is consistent with the absence of a strong Nernst signal well
above T$_c$ \cite{balci:054520} and the absence of an extended
pseudogap phase \cite{daganQazilbashtunneling} in electron-doped
cuprates.
\par

Next we calculate the BCS parameters inferred from our data (see
Table \ref{myTable}). First, we find $2\Delta_{max}/kT_c$ for the
various doping levels, where the low temperature values of
$\Delta_{max}$ are used. Within experimental errors, our results
yield the BCS weak-coupling ratio of 3.5. Unfortunately, there is
no available theoretical estimation of $2\Delta_{max}/kT_c$ for
the nm$d$ case. Second, we calculate $\xi_{BCS}=\frac{\hbar
v_{_F}}{\pi\Delta_{max}}$ where the averaged Fermi velocity $v_F$
can be estimated from photoemission measurements to be 3.75$\cdot
10^5$ m/sec.\cite{armitage:064517}. Using H$_{c2}$ evaluated
earlier, we calculate the Ginzburg-Landau coherence length,
$\xi_{GL}=\sqrt{\frac{\phi_0}{2\pi H_{c2}}}$, with $\phi_0$ the
flux quantum. Our results are in agreement with Raman spectroscopy
data \cite{qazilbashRamanSCstate}.
\par
Despite the fact that hole-doped cuprates are considered to be
``bad metals'' in their normal state, they are still in the clean
limit. This is due to their short coherence length
\cite{YayuWangHC2fromnernstandxifromarpes}. Here, we find a BCS
coherence length an order of magnitude larger than in the
hole-doped case. Therefore, it is possible that electron-doped
cuprates are closer to the dirty limit, \emph{i.e.}
$\xi_{BCS}>\ell$, where $\ell$ is the mean free path. Since
$\xi_{GL}/\xi_{BCS}$ is well below its clean limit value of 0.74,
we roughly estimate (see Table \ref{myTable}) $\ell$ using the
relation for the dirty limit $\xi_{GL}=0.855(\xi_{BCS}\ell)^{1/2}$
\cite{Tinkham}. Indeed, we find that we are in the dirty limit.
Recently, Homes \emph{et.al.} came to similar conclusions from
optics measurements on Pr$_{1.85}$Ce$_{0.15}$CuO$_{4-\delta}$
samples \cite{homes:214515}.
\par
\par

We can independently estimate $\ell$ from resistivity using
$k_F\ell=\frac{hd\sigma_{ab}}{e^2}$ to verify our results. Here, d
is the distance between two successive CuO$_2$ planes,
$\sigma_{ab}$ is the in-plane conductivity, and $e$ is the
electron charge. First, we notice that the doping dependence of
$\ell$ obtained from tunneling (within the error bars) is
consistent with that of the resistivity
\cite{daganResistivityPRL}. Second, we can obtain an upper limit
for $\ell$ by estimating the Fermi wave number, $k_F\simeq
\frac{mv_{_F}}{\hbar}$ using $m$ as the bare electron mass and
$v_F$ from photoemission measurements \cite{armitage:064517}. For
x=0.19 we obtain from the resistivity at 0.3 K, $\ell=23$nm, in
excellent agreement with the tunneling estimation.
\par
Our tunneling spectra are very different from those obtained on
hole-doped YBa$_2$Cu$_3$O$_7$ (YBCO) using the same counter
electrode (Pb) \cite{Lesueur}. First, while the data in Fig
\ref{MainResults} exhibits conductance peaks at
$\Delta(PCCO)\pm\Delta(Pb)$ as in classical tunneling experiments
\cite{GiaeverRMP}, for the case of in-plane tunneling into YBCO
the peaks appear at $\Delta(YBCO)$ and $\Delta(Pb)$ separately.
Moreover, strong in-plane anisotropy is observed for various film
orientations in YBCO in contrast to PCCO \cite{shan:144506}.
\par
Although our experiment alone can not exclude a possible
anisotropic $s$-wave order-parameter, the tunneling fits are
consistent with phase sensitive experiments for optimum doping
\cite{TsueiElectronDoped} and slight over-doping \cite{Hilgenkamp}
that indicate the existence of nodes and phase changes in the
order-parameter. However, $d$-wave or $n$md order-parameters
should result in a zero-bias conductance peak for in-plane
tunneling spectroscopy. This zero-bias conductance peak has never
been observed in low transparency junctions ($Z>1$) in contrast
with hole-doped cuprates where it can be hardly avoided
\cite{deutscherRMP,DaganEPL}. Another puzzle is lack of
directional in-plane sensitivity in tunneling measurements
\cite{shan:144506}. These discrepancies between quasiparticle
tunneling \cite{shan:144506} and Cooper pair tunneling
\cite{TsueiElectronDoped,Hilgenkamp} may be resolved assuming
dirty superconductivity as we report. For the first case,
quasiparticles undergo several scattering events within a
coherence length and thus become insensitive to their initial
momentum and phase. Therefore, the initial direction from which
the quasiparticles are injected is unimportant. By contrast,
Cooper-pairs retain their phase information over a coherence
length scale and consequently experiments involving their
tunneling can probe the anisotropic phase of the order-parameter.
However, it still deserves theoretical attention to resolve the
question how an order-parameter which changes sign, such as nm$d$,
can still survive in the dirty limit.  For such an order-parameter
any scattering center is a pair breaker \cite{VarmaSachdevMillis}.

\par

In summary, we measured tunneling conductance for
Pb/Insulator/Pr$_{2-x}$Ce$_x$CuO$_{4-\delta}$ over wide doping,
temperature and magnetic field ranges. Taking advantage of the
accessible upper critical field of the electron-doped cuprates,
the well studied superconducting parameters of lead, and the clear
features in S/I/S contacts, we were able to properly normalize and
fit the data even at relatively high temperatures. The data fits a
non-monotonic $d$-wave behavior (for PCCO) over the entire doping
range. The gap maximum follows the BCS temperature dependence for
underdoped and optimally doped samples. The upper critical field
extracted from our tunneling data and the gap amplitude are used
to calculate the BCS and the Ginzburg-Landau coherence lengths.
From these length scales we estimate the mean free path and
conclude that PCCO is in the dirty limit. This may explain the
absence of a zero bias conductance peak in low transparency
junctions and the lack of in-plane directional sensitivity in the
tunneling spectrum.

\begin{acknowledgments}
We thank G. Deutscher and G. Blumberg for useful discussions and
A. Chubukov for many discussions and for suggesting the nm$d$
fitting function. Support from NSF grant number DMR 0352735 is
acknowledged. R. B. acknowledges support from Israeli Science
Foundation and the Heinrich-Hertz Minerva Center for High
Temperature Superconductivity. Y. D. acknowledges support from the
German-Israeli Foundation.
\end{acknowledgments}

\bibliographystyle{apsrev}

\bibliography{PCCOTunneling}

\end{document}